\documentclass[twocolumn,showpacs,preprintnumbers,amsmath,amssymb]{revtex4}

\usepackage{graphicx}
\usepackage{dcolumn}
\usepackage{bm}

\begin{document}


\title{Theory of spatial mode competition in a fiber amplifier}

\author{Anatoly P. Napartovich}
 
\author{Dmitry V. Vysotsky}%
 \email{dima@triniti.ru}
\affiliation{%
SRC RF Troitsk Institute for Innovation and Fusion Research, Troitsk, Moscow region, 142190, Russia
}%

\date{\today}

\begin{abstract}
Theory of monochromatic wave field amplification in a waveguide array based on expansion of the wave field in 
terms of guided array modes is developed. Equations for the expansion coefficients include cross-modal gain,
which completely changes behavior of the amplified wave field. Analysis of two-mode amplification reveals 
new unusual features in characteristics of two-mode amplifier. Instead of unlimited growth of both modes 
for incoherent fields, one of the modes grows with no limit and suppresses the lower-power mode. Effects 
associated with the cross-modal gain are illustrated analytically on a system of two thin  
parallel planar waveguides. Conditions are found when a mode with lower gain can become the dominant one 
at the output of amplifier.
\end{abstract}

\maketitle

\section{\label{sec1}Introduction}
Wave field propagation over an array of parallel waveguides is up to now a subject of numerous 
theoretical and experimental studies, mostly for optoelectronic devices. In particular, non-linear 
refraction of materials can lead to a wide variety of non-trivial physical phenomena in the 
waveguide arrays, thus making them very appealing for research. Lattice solitons \cite{latsol} 
and non-linear optical switches \cite{switches} are among such phenomena. 
An array of nonlinear waveguides in a multi-core fiber (MCF) laser \cite{Glas} 
is another example very important for the laser design. The MCF laser construction is an array 
of single-mode cores doped by rare-earth ions in a common pump cladding. Such a design allows 
one to enlarge a combined mode area, thus diminishing the fiber length. Single-mode cores can 
be arranged also in a form of active defects in photonic crystal fiber lasers \cite{PCF}. 
The MCF design has an advantage over single-core large mode area constructions only if the fields in the cores are phase locked. Phase locking can be provided by an external optics or by diffractive exchange of fields between the cores. 

The purpose of this article is to analyze the mode competition in the multi-core fiber amplifier. 
This problem seems to be important far outside the fiber laser theory because it can be considered 
as a general dynamic problem for the Schroedinger equation with a complex-valued nonlinear potential. 
The nearest quantum mechanical analogy for the optical wave field propagation is wave function temporal 
variation in an array of potential wells.   

The method commonly used in describing the guided field evolution in passive systems is a coupled mode theory \cite{Marcuse} (CMT), in which the wave field is presented as a sum of sole waveguides modes with unknown 
coefficients being a function of the propagation distance. These coefficients satisfy equations, which are 
derived by perturbation method. CMT is intuitively evident and can be easily extended to new physical effects 
\cite{Yariv}. The CMT for a pair of nonlinear waveguides was analyzed in \cite{Yensen, Majer}. 
This theory was generalized by Hardy and Streifer \cite{Hardy} who introduced non-identical waveguides and vector field effects. Phase self-synchronization at high intensity $\left(\sim 1~\text{GW/cm}^2\right)$ and interaction of
wave fields with different polarizations and frequencies in systems of 2 and 3 evanescently coupled waveguides
were topics of earlier work, which was summarized in \cite{switches}. 

Recently \cite{Peles}, the CMT approach was implemented for analysis of dynamic stability for MCF configurations with a separate multi-channel coupler. This work was inspired by experiments  \cite{Monica}. Using model 
coupling matrix, the authors  \cite{Peles} have explained the robustness of the phase synchronization 
in such systems provided special asymmetry is introduced into the construction. 

An MCF laser with 7-core hexagonal structure has exhibited experimentally \cite{CheoExp} the far field 
patterns typical for phase-locked operation at power level more than 100 W. 
An analogous 19-core fiber amplifier \cite{OptExpr} has achieved 20 dB gain at near to diffraction limit beam
quality. The calculation based on the CMT predicts \cite{BochoveOpL} that the observed self-synchronization 
stems from intensity dependence of the refractive index of rare-earth doped fiber. However, direct numerical 
modeling based on 3D beam propagation method \cite{OpComm} (BPM) 
has clearly shown \cite{KvElectr}, that the gain non-uniformity in the system is the major factor responsible for the effect, while the role of refractive index nonlinearity is subsidiary.   

An important feature of the MCF amplifier is that the field distribution in a transverse plane is strictly determined by the index profile, while distortions induced by gain being negligible. This fact approves an approach based on wave field expansion over passive structure modes, which can be found easily by a standard solver. In particular, such modes for the 7-core fiber design were found in \cite{CheoJosab} by a finite element solver.

Comparison between results of 3D BPM calculations  \cite{KvElectr} and of modal analysis revealed \cite{Lghtw} 
a seeming contradiction: an in-phase mode according to mode solver possesses the lowest gain, while 
according to 3D BPM approach the wave field converges to this mode, if the differences in phases of 
launched beams are within a few tenth of radian. To resolve this contradiction, the theory of light 
amplification in the system of parallel waveguides with saturable gain should be reconsidered. 

Mode competition for gain has been studied since early times of laser researches. Differing spatial profiles of modal intensities result in gain spatial hole burning effect, which, in turn, induces instability of single mode lasing \cite{Rautian, SatzTang}. Earlier studies
concentrated on the competition of optical modes with different wavelengths. An analysis of oscillation 
dynamics in a two-mode laser cavity was performed in \cite{Oraevski, Fleck}. It is traditionally believed that this dynamics is described satisfactorily taking into account an effect of the modal gain saturation by 
the intensity of another mode (so-called cross-saturation, it was studied mainly for counter propagating 
modes of ring lasers \cite{MandelWlf}). 

It is shown below that the major factor governing competition of the modes in the 
multi-core amplifier is a cross-gain \cite{Cleo}, which is defined as the product of the mode 
field profiles with the gain distribution integrated over fiber cross section area. General 
equations governing wave field propagation over the MCF amplifier are formulated in section \ref{sec2}. The mode 
suppression mechanism in two coupled waveguides is discussed in section \ref{sec3}. A system of 2 delta-function 
type waveguides is considered in sections \ref{sec4}, where optical modes are found, and \ref{sec5}, where simultaneous 
amplification of two modes in this system is analyzed. Amplifier characteristics are compared for cases 
of two incoherent and two coherent modes. A weak mode suppression mechanism by a strong one is identified for two-delta-function-type waveguides. The conclusion is made that this mechanism associated with cross gain, 
which remains unnoticed till the last time, is expected to exist in any system of active waveguides.

\section{\label{sec2}Basic equations}
The  built-in index profile of a  multicore fiber amplifier is step-wise function with higher index regions, 
called cores, each guide single mode. To approach the goal to have the large mode area, small values  
$\left(\sim 10^{-3}\right)$ of the core-cladding index difference $\Delta n$ are used in experiments. Besides, we do not consider 
birefrigence effects. It is then possible \cite{Snyder} to use the scalar approximation instead of soving 
Maxwell's equations for the elecromagnetic field in the fibre. The wavefield can be characterized in this 
approximation by a scalar function $\psi(x,y)\exp(i\beta z-i\omega t),$ $z$ is the propagation distance,  $\omega=kc$ is the oscillation frequency, $k$ is the vacuum wave number, $c$ is the speed of light, $\beta$ is the propagation 
constant.  The function $\psi$  could be any transverse component of electrical or magnetic field, obeying 2D 
Helmholtz equation, which formally coincides with stationary Schroedinger equation  in quantum mechanics: 
\begin{equation}
{\rm\bf H}\psi=-\beta^2\psi,~~{\rm\bf H}\psi=
-\left(\frac{\partial^2\psi}{\partial x^2}+\frac{\partial^2\psi}{\partial y^2}\right)+V\psi,
\label{Schroed}
\end{equation}  
where  $\rm\bf H$ is Hamiltonian,  $\beta^2$ plays the role of energy, $x$ and $y$ are the transverse spatial 
variables, and the potential function $V=-k^2n^2(x,y),$  $n$ is the  refractive index. Equation (\ref{Schroed}) is solved 
with corresponding conditions at the core-cladding boundary and at the infinity. This problem has a continuous 
spectrum of the radiating modes $\psi_Q$ and a discrete spectrum of the guided modes $\psi_j$, so any field 
injected into the MCF can be presented as the sum over the spectrum of $\psi_j$ plus the integral over the spectrum of  $\psi_Q$ with some  coefficients \cite{Snyder}.
The field amplitudes of guided modes $\psi_j(x,y)$ are real-valued functions for  no-loss fiber. Besides, 
the guided and radiating modes are mutually orthogonal between each other.

The wave field $\Psi(x,y,z)$ in active fiber with gain $g\left(x,y,\Psi\right)$ satisfies the 3D Helmholtz equation:
\begin{equation}
\frac{\partial^2\Psi}{\partial z^2}+\frac{\partial^2\Psi}{\partial x^2}+
\frac{\partial^2\Psi}{\partial y^2}+\left(k^2n^2-ikn_0g\right)\Psi=0, 
\label{eiggen}
\end{equation}
 $n_0$ is the cladding index. Thus, in contrast to quantum mechanics the potential in laser optics is a 
complex-valued function $V+ ikn_0g\left(x,y,\Psi\right).$

Being interested in behavior of guided modes in the amplifier, it is convenient to take the modes of equation 
(\ref{Schroed}) as a basis for expansion of the wave field: 
\begin{equation}
\Psi(x,y,z)=\sum_j c_j(z)\psi_je^{i\beta_jz}+\int c_Q(z)\psi_Qe^{i\beta_jz}dQ.
\label{expns}
\end{equation}
The non-uniform gain leads to an interaction between guided 
and radiating (leaky) modes, thus producing additional losses for the guided modes due to carrying away the energy by 
the radiating modes. However, transformation of guided modes into 
leaky ones is a negligible effect for typical parameters of fiber amplifiers, as it was confirmed by 3D BPM 
calculations \cite{LectNot}. From the other side, leaky modes see the gain much lower than guided modes because
the cladding area is much larger than the area of active cores.  For these reasons we can substituting expansion 
(\ref{expns}) into (\ref{eiggen}) and neglect the radiating modes. The gain in the cores is usually of the order 
of $10^{-1}$ $\text{cm}^{-1},$ while a typical spectral separation between the propagation constants of guided modes is $\sim 10$ $\text{cm}^{-1}$. Thus, variation of the guided modes amplitudes $c_j$ due to amplification 
is slow process and it is possible to use approximation, in which the terms $d^2c_j/dz^2$ are neglected. 
Then the following system of equations for the expansion coefficients  $c_j$  can be derived by 
multiplying the resulted equation to $\psi_j\exp(-i\beta_jz)$ and integrating over the fiber aperture:   
\begin{equation}
\frac {dc_j}{dz}=\frac 12c_jg_{jj}+\frac 12\sum_{l\neq j}g_{jl}c_le^{i(\beta_l-\beta_j)z},
\label{coef0}
\end{equation}
where summation is made over {\it N} guided modes. The optical modes introduced are normalized for convenience as $\iint \psi_j^2\,dxdy=1.$ The difference of propagation constants is small parameter in comparison with $kn_0,$  so the factors $\beta_j/\left(kn_0\right)$ in (\ref{coef0})  have been replaced by one. The applicability of this approximation was verified by 3D modelling of the MCF aplifier.  Matrix elements  $g_{jl}$ describe an interaction between the wave field and the gain:
\begin{equation}
g_{jl}=\iint g(x,y)\psi_j\psi_l\, dxdy,
\label{ingrl}
\end{equation} 
where integration is made over the fiber aperture. The coefficients $g_{jj}$ are the modal gains of $j$-th mode, 
while the coefficients $g_{jl}$ in the sum in (\ref{coef0}) describe the cross-modal gain effect.    

System (\ref{coef0}) of complex equations can serve as the basis for analyzing the mode competition in a wide variety of
MCF amplifiers. To give an idea about new features, which possesses the theory based on equation system (\ref{coef0}),
let us consider the simplest situation when a waveguide array supports only two guided modes. In this case, system (\ref{coef0}) can be reduced to 3 ordinary equations for real-valued functions
\begin{subequations}
\label{eqsX}
\begin{eqnarray}
\frac {dP_1}{dz}=g_{11}P_1+\sqrt{P_1P_2}g_{12}\cos\phi,
\label{eq1}
\\
\frac {dP_2}{dz}=g_{22}P_2+\sqrt{P_1P_2}g_{12}\cos\phi,
\label{eq2}
\\
\frac {d\phi}{dz}=\delta\beta-\frac{P_1+P_2}{\sqrt{P_1P_2}}g_{12}\sin\phi.
\label{eq3}
\end{eqnarray}
\end{subequations}
Here the modal powers $P_j,$ $j=1,2$ are introduced by an expression $c_j=\sqrt{P_j}\exp(i\phi_j-i\beta_jz),$
$\phi=\phi_2-\phi_1$ is the phase difference between the modal fields, $\delta\beta=\beta_2-\beta_1$.
$P_j$  is a fraction of the wave field power carried by the $j$-th mode. Equation (\ref{eq3}) describes
the behavior of the modal phase difference.  Since gain coefficients are much smaller than  $\delta\beta$ 
value, this phase difference grows along the propagation axis practically linearly, 
$d\phi/dz\approx\delta\beta.$ In this approximation, the problem is reduced to solving equations
(\ref{eq1}) and (\ref{eq2}), only.
 
Spatial beatings between two modes lead to oscillatory modulation of the total field intensity along axis with a characteristic spatial frequency of a few 10 $\text{cm}^{\text{-1}}.$ This modulation induces through
the gain saturation similar oscillations in the material gain. The resulting non-uniform gain profiles 
in different cores correlate with modal powers. In this case we have a tangled interaction between 
the wave field and transverse and longitudinal gain non-uniformities. Our purpose is to make this 
interaction clearer using simplest waveguide configuration.

It should be noted, that dynamic equations for the two-mode laser \cite{Oraevski} have this sort of 
mathematical structure as eqs. (\ref{eqsX}). However, the physical content of the problem studied is 
essentially different: the authors \cite{Oraevski} analyzed stability of two-mode regimes in a 
two-wavelength laser, while we study non-linear effects in an amplifier with continuous wave 
(cw) input signal.

\section{\label{sec3}Two mode competition in the MCF}
To complete the system of equations (\ref{eqsX}) it is necessary to specify how the material gain 
coefficient depends on the total intensity. Generally, kinetics of population inversion varies 
from one type of laser to another. The common feature for all cw systems is so-called gain saturation, 
i.e. reduction of inversion and gain induced by stimulated emission. This effect can be qualitatively 
well described by the simplest formula $g=g_0/\left(1+I/I_{\text{Sat}}\right),$ where $g_0(x,y)$ is 
the small signal gain, $I$ is the wave field intensity, $I_{\text{Sat}}$ is the saturation intensity.
We will normalize the intensity to the saturation intensity value. 

Even with such a simple model for gain saturation and taking the mode phase difference as 
$\phi\approx\delta\beta\cdot z,$ the system (\ref{eqsX}) is still rather difficult for an analysis. The point 
is that $g_{jl}$ terms defined by (\ref{ingrl}) are complicated functions of $P_{1,2},$ which cannot 
be found explicitly for realistic waveguide structures. Nevertheless, some general properties of solutions 
to (\ref{eqsX}) can be identified.

An important distinction of these equations is the identity of the second terms in right-hand side of equations
 (\ref{eq1}) and (\ref{eq2}). For this reason it can be deduced from eqs. (\ref{eq1})--(\ref{eq2}) for the 
evolution of the mode power ratio:
\begin{equation}
\frac d{dz}\left(\frac{P_2}{P_1}\right)=\frac{P_2}{P_1}\left(g_{22}-g_{11}+
g_{12}\cos\phi\cdot\frac{P_1-P_2}{\sqrt{P_1P_2}}\right).
\label{rat}
\end{equation}
If  $g_{12}\cos\phi$  is negative, then the last term in bracket of the right-hand side of (\ref{rat}) 
supports the trend the power ratio to grow when this ratio is greater than 1.
 
The dimensionless total wave field intensity can be expressed in a form
$I=P_1\psi_1^2+P_2\psi_2^2+2\sqrt{P_1P_2}\psi_1\psi_2\cos\phi.$ The cross gain variation as a function of $z$ can be understood from an expression
\[g_{12}\cos\phi=\iint \frac{g_0\psi_1\psi_2\cos\phi\,dxdy}{1+P_1\psi_1^2+P_2\psi_2^2+2\sqrt{P_1P_2}\cos\phi},
\]
which is obtained from (\ref{ingrl}). The mode amplitude product necessarily changes sign within 
the cores to provide orthogonality of modes. It is seen
that the integrand generally takes on negative value with larger absolute magnitude when 
$\psi_1\psi_2\cos\phi<0.$ Therefore the quantity $g_{12}\cos\phi$ is preferably negative.
It can be rigorously proved that  $g_{12}\cos\phi$ is non-positive for a system of two parallel waveguides 
possessing mirror symmetry $\Delta n(x,y)=\Delta n(-x,y)$ and supporting two guided modes.
One of modes is symmetric  $\left(j=1\right)$ and another mode is antisymmetric  
$\left(j=2\right).$ 
Taking into account symmetry properties, the cross gain coefficient can be expressed as
\begin{equation}
g_{12}\cos\phi=-4\sqrt{P_1P_2}\cos^2\phi\iint_{S_1}\frac{g_0\psi_1^2\psi_2^2}{C_{\psi}}\,dxdy,
\label{ga12}
\end{equation}
where the integration is made over one of the waveguides, and the saturation factor $C_{\psi}$ is 
\[
C_{\psi}=\left(1+P_1\psi_1^2+P_2\psi_2^2\right)^2-4P_1P_2\psi_1^2\psi_2\cos^2\phi.\]
It is evident from (\ref{ga12}) that the term $g_{12}\cos\phi$ in this case is always non-positive. 
It turns to zero at $P_1=0$ or $P_2=0.$

Since the cross-gain term $g_{12}\cos\phi$ equally diminishes an amount of energy extracted by stimulated emission for 
both modes, the increase of the mode power is favored for the mode with higher power. This is rather 
important conclusion radically differing from the intuitive suggestion  that the mode possessing higher saturated gain is dominant at an output of a sufficiently long amplifier. It is shown below, that the mode can possess
higher modal gain throughout the entire amplifier but the part of the power carried with this mode in the total power
diminishes in a course of amplification.

The idea of using the saturated modal gains \cite{Moloney} stems from consideration of the incoherent fields of competing modes. 
Such a situation can be thought of as the competition of two signals launched in the fiber 
amplifier from independent sources at the same frequency. The wave field intensity is expressed 
in the incoherent case as  $I=P_1\psi_1^2+P_2\psi_2^2,$ 
and the equations for the two-mode evolution are 
\begin{equation}
\frac{dP_1}{dz}=g_{11}P_1,\qquad\frac{dP_2}{dz}=g_{22}P_2,
\label{incoh}
\end{equation}
$g_{11}$ and $g_{22}$  are the modal gains of modes 1 and 2, correspondingly.
These equations include so-called cross saturation of gain associated with the fact that 
both modes saturate gain. In the limit of a weak saturation, equations (\ref{incoh}) 
are reduced to the form closely resembling known dynamic equations for two-mode ring laser
(\cite{Lamb})
\begin{equation}
\begin{array}{rcl}
\dot{X}&=&2X\left(\alpha_1-\beta_1X-\theta_{12}Y\right),\\
\dot{Y}&=&2Y\left(\alpha_2-\beta_2Y-\theta_{21}X\right),
\end{array}
\label{LMB}
\end{equation}
where $X,$ $Y$ are dimensionless intensities of  modes 1 and 2, $\alpha_j,$ and $\beta_j,$ $j=1,2$ are the 
above-threshold small signal gains and self saturation coefficients, respectively. $\theta_{12}$ and $\theta_{21}$  
are the cross saturation coefficients.  If there is a difference in the modal gains, then one mode may suppress 
the growth of the other. If the modal gains are equal, both the modes lase the same power for the case of an inhomogeneously broadened $\left(\theta_{12}\theta_{21}\leq \beta_1\beta_2\right)$ 
ring laser \cite{Lamb}. For a homogeneously  broadened $\left(\theta_{12}\theta_{21}>\beta_1\beta_2\right)$ ring laser \cite{Singh} equations (\ref{LMB})
predict that from two modes starting at $t=0$ the mode with higher power suppress completely the second mode. 
Practically, the operation regime of a laser in the last case is random due to the spontaneous emission effects 
\cite{SinghReport, MandelWlf}.

Actually, the analogy between weak-saturation limit of (\ref{incoh}) and laser equations 
(\ref{LMB}) is valid only for low-power wave fields, i. e. for laser starting from small signal.

\section{\label{sec4}System of two ultra-thin planar waveguides}
The delta-function $\left(\delta(x)\right)$ is a favorite potential well in quantum mechanics. 
In application to optics, delta-function well is a mathematical limit of the high-contrast 
thin planar waveguide with arbitrarily small waveguide width  $d$ and a high refractive index difference 
$\Delta n$ characterized by a single parameter $d\Delta n.$ In this limit, the waveguide supports single mode, 
the wave field of which is almost constant within the core and extends far outside the core. It should be mentioned, that the core cladding difference is restricted by the condition $\Delta n\ll 1$ for the scalar model (\ref{Schroed}) applicability. Usage of the delta-function as a model waveguide allows one to study waveguide arrays analytically \cite{Kivshar}. 

\begin{figure*}
\includegraphics{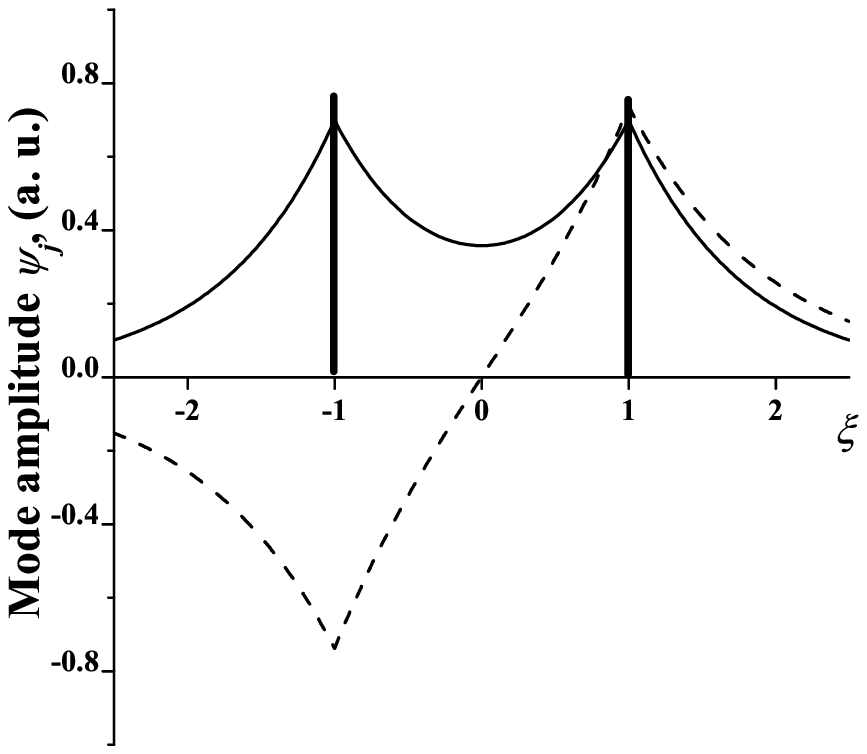}
\caption{\label{fig:sch} Schematic of the system of two parallel ultra-thin waveguides. The profiles of the symmetric (solid line) and antisymmetric (dashed line) modes correspond to $\kappa^{\prime}=1.2.$ }
\end{figure*}

\begin{figure*}
\includegraphics{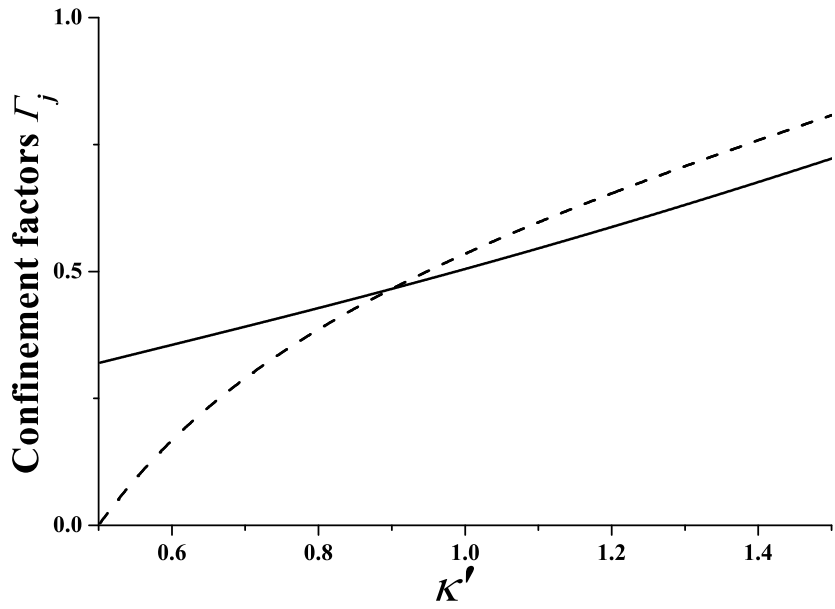}
\caption{\label{fig:gamm} Confinement factors for the symmetric (solid line) and antisymmetric (dashed) modes vs. the coupling strength parameter  $\kappa^{\prime}$. The quantities are dimensionless.}
\end{figure*}

Figure \ref{fig:sch} shows the system of two ultra-thin waveguides situated at locations  $x=\pm a.$ For this particular system
it is convenient to introduce dimensionless transverse coordinate $\xi=x/a.$ Equation (\ref{Schroed}) for guided modes
 of the waveguide array reads
\begin{equation}
\frac{d^2\psi}{d\xi^2}+\left\{-\alpha^2+2\kappa\left[
\delta\left(\xi-1\right)+\delta\left(\xi+1\right)\right]\right\}\psi=0,
\label{dlt}
\end{equation}
where $\kappa=k^2adn_0\Delta n$ is the delta-function amplitude, $\alpha$ is the wave field attenuation rate 
outside of the waveguides,  $\alpha^2$ is eigenvalue characteristic for a given mode. 
As well known \cite{Kogan}, such a system supports two guided modes shown in Fig. \ref{fig:sch}, provided the 
condition $2\kappa \geq 1$ is fulfilled. The fundamental mode is symmetric $\psi_S\left(\xi\right)$ 
(in-phase mode), and the second mode is antisymmetric, $\psi_A\left(\xi\right)$ (out-of-phase mode). The amplitudes of these modes attenuate exponentially at $\xi\rightarrow\pm\infty$ with corresponding rates $\alpha_S$ and $\alpha_A$, while in the space between the waveguides $\psi_S\sim\cosh(\alpha_S\xi),$ and $\psi_A\sim\sinh(\alpha_A\xi).$ 
The attenuation rates satisfy transcendental equations 
\begin{equation}
\alpha_S=\kappa\left(1+e^{-2\alpha_S}\right),
\quad\alpha_A=\kappa\left(1-e^{-2\alpha_A}\right).
\label{dsprsn}
\end{equation}
The modal shift of propagation constant $\delta\beta_j=\beta_j-kn_0$ is expressed as 
$\delta\beta_j=\alpha_j^2/\left(4L_R\right),$ $L_R=kn_0a^2/2$ is the Rayleigh length. 
Gain in the waveguides can be included in the model by addition an imaginary part into   
$\kappa$: $\kappa=\kappa^\prime+i\kappa^{\prime\prime},$ where 
$\kappa^{\prime\prime}=g\kappa^{\prime}/(2k\Delta n).$ The modal gains in the small signal limit read
\begin{eqnarray*}
G_S=\frac{2\alpha^{\prime}_S\kappa^{\prime\prime}}{L_R}
\left[1+\frac{\sinh 2\alpha^{\prime}_S+
\left(\sin 2\alpha^{\prime\prime}_S\right)\alpha^{\prime}_S/\alpha^{\prime\prime}_S}{\sinh 2\alpha^{\prime}_S+
\cos 2\alpha^{\prime\prime}_S}\right]^{-1},\\
G_A=\frac{2\alpha^{\prime}_A\kappa^{\prime\prime}}{L_R}
\left[1+\frac{\sinh 2\alpha^{\prime}_A-
\left(\sin 2\alpha^{\prime\prime}_A\right)\alpha^{\prime}_A/\alpha_A^{\prime\prime}}{\sinh 2\alpha^{\prime}_A-
\cos 2\alpha^{\prime\prime}_A}\right]^{-1},
\end{eqnarray*}
where $\alpha^{\prime\prime}_j$ and $\alpha^{\prime\prime}_j$ are the real and imaginary parts of the corresponding parameters satisfying (\ref{dsprsn}).
For the symmetric system under consideration it is convenient to introduce a confinement factor 
of the mode  $\Gamma_j,$ as being proportional to the overlap of the mode intensity and gain in \textit{one} 
waveguide. For the ultra thin waveguides the confinement factor is equal to the squared amplitude of the mode in the waveguide. The modal gain is expressed as $G_j=2\Gamma_jgd/a.$  As far as $\kappa^{\prime}\gg\kappa^{\prime\prime}$ $\left(2k\Delta n\gg g\right),$ 
the expressions for modal gains and confinement factors can be simplified. The confinement factors of the symmetric and antisymmetric modes can be expressed as 
\begin{eqnarray*}
\Gamma_S=\frac {\kappa^{\prime}}2\cdot\frac{\left[1+\exp\left(-2\alpha_S^{\prime}\right)\right]^2}
{1+2\kappa^{\prime}\exp\left(-2\alpha_S^{\prime}\right)},\\
\Gamma_A=\frac {\kappa^{\prime}}2\cdot\frac{\left[1-\exp\left(-2\alpha_A^{\prime}\right)\right)^2}
{1-2\kappa^{\prime}\exp\left(-2\alpha_A^{\prime}\right)}.
\end{eqnarray*}
$\alpha_j$ $(j=S,A)$ satisfy equations (\ref{dsprsn}) with $\kappa$ replaced by $\kappa^{\prime}$. 
In the limit of weak coupling between waveguides $\left(\kappa^{\prime}\gg 1\right)$ the antisymmetric mode 
has higher confinement factor, than the symmetric one:
\begin{eqnarray*}
\Gamma_S\approx\frac{\kappa^{\prime}}2
\left[1+2\kappa^{\prime}\exp\left(-2\alpha_S^{\prime}\right)\right]^{-1},\\
 \Gamma_A\approx\frac{\kappa^{\prime}}2\left[1-2\kappa^{\prime}\exp\left(-2\alpha_A^{\prime}\right)\right]^{-1}.
\end{eqnarray*}
As coupling between waveguides increased the antisymmetric mode tends to transform into a leaky mode, 
$\left(\alpha_A\rightarrow 0\right),$ and its modal gain diminishes proportional to  $\alpha_A,$ while the symmetric mode gain remains to be of finite value. Thus, there is a critical distance between the waveguides, at which both gains are equalized. The confinement factors are shown in Fig. \ref{fig:gamm} as functions of   
$\kappa^{\prime}.$ The curves intersect at 
$\kappa^{\prime}_{\text{ cr}}=0.900126\dots$. This value corresponds to $a\approx 3.9\,\mu\text{m}$ for the system 
of two waveguides with $\Delta n=2\cdot 10^{-3},$ $d=2\,\mu\text{m},$ $n_0=1.456$ and the radiation wavelength 
$2\pi/k=1~\mu\text{m}.$ The Rayleigh length $L_R$ in this case is 70 $\mu\text{m}.$
 
\section{\label{sec5}Wave field amplification in the system of two ultra-thin waveguides}
It is instructive to analyze wave field amplification in two $\delta$-function type coupled waveguides taking as a reference case the standard equations applicable for description of two incoherent modes with gain cross saturation accounted. For definiteness, the simplest gain saturation model is adopted
$g=g_0/\left(1+I/I_{\text{Sat}}\right),$  and field intensity in following is measured in $I_{\text{Sat}}$ units.

\subsection{Incoherent wave fields}
\begin{figure*}
\includegraphics{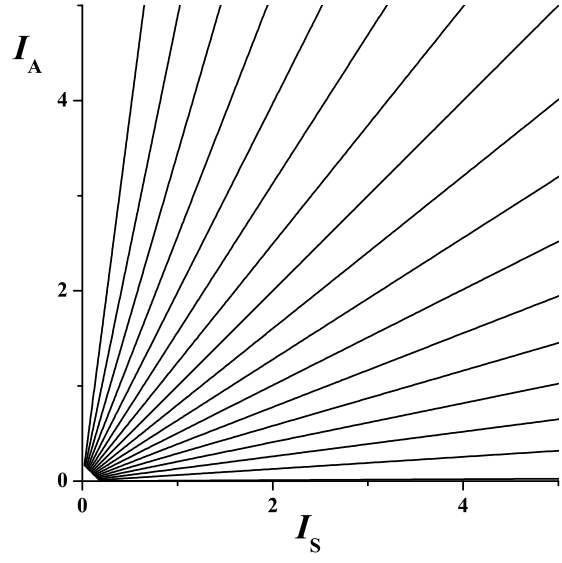}
\caption{\label{fig:inch}Mode intensities in the waveguide normalized to $I_{\text{Sat}}$ for different values of initial powers ratio for the incoherent wave fields of the modes,  $\Gamma_S=\Gamma_A.$}
\end{figure*}

For the incoherent fields of two modes in the waveguide system shown in Fig. \ref{fig:sch} 
the equations for the modal powers (\ref{incoh}) read 
\begin{eqnarray*}
\frac{dP_S}{d\zeta}=\frac{2\kappa_0^{\prime\prime}P_S\Gamma_S}{1+P_S\Gamma_S+P_A\Gamma_A},
\\
\frac{dP_A}{d\zeta}=\frac{2\kappa_0^{\prime\prime}P_A\Gamma_A}{1+P_S\Gamma_S+P_A\Gamma_A},
\end{eqnarray*}
where $\zeta=z/L_R$ and $\kappa_0^{\prime\prime}=g_0\left(L_Rd/a\right).$ 
This system of equations can be easily integrated 
\[
\ln P_S/P_{S0}+\Gamma_S\left(P_S-P_{S0}\right)+\Gamma_S\left(P_A-P_{A0}\right)=2\Gamma_S\kappa_0^{\prime\prime}\zeta,
\]
where $P_A/P_{A0}=\left(P_S/P_{S0}\right)^{\gamma},$ $\gamma=\Gamma_A/\Gamma_S,$ $P_{S0}$ and $P_{A0}$ are 
the mode powers at the amplifier entrance. An asymptotic $\left(\zeta\rightarrow\infty\right)$ behavior 
of the modes depends on a value of $\gamma,$ which is equal to ratio of modal gains. Namely, at $\gamma=1$
both modes grow with equal rates. This case is illustrated in Fig. \ref{fig:inch}, which shows the diagram in dimensionless variables $I_S=P_S\Gamma_S$ and $I_A=P_A\Gamma_A.$ It is clearly seen that the proportion $P_A/P_S$
remains constant at any distance. If the modes have different overlaps with the gain, the mode with higher confinement factor wins and its power increases linearly with length, while the power of the second mode grows as a fractional power of the length. It means that powers of both modes increase with no limit. This result does not depend on the initial proportion of the powers of the modes, if the amplification is large enough. 

\subsection{\label{sec5:level2}Coherent wave fields}
If the wave fields of two competing modes are coherent, gain coefficients entering into the system 
of equations (\ref{eqsX}) can be found explicitly from eqns. (\ref{ingrl}) and (\ref{ga12}) as
\begin{equation}
\begin{array}{rcl}
g_{jj}&=&2\kappa_0^{\prime\prime}\Gamma_j\left(1+P_S\Gamma_S+P_A\Gamma_A\right)/\left(CL_R\right),\\
g_{AS}&=&-4\kappa_0^{\prime\prime}\Gamma_A\Gamma_S\sqrt{P_AP_S}\cos^2\phi/\left(CL_R\right),
\end{array}
\label{coefcalc}
\end{equation}
here $j=A,S$ and the denominator contains 
\[C=\left(1+P_S\Gamma_S+P_A\Gamma_A\right)^2-4\Gamma_A\Gamma_SP_AP_S\cos^2\phi.\]
In the specific system under consideration the ratio of the modal gains of the two modes is independent from the field
in the amplifier:
\[g_{SS}/g_{AA}=\Gamma_S/\Gamma_A.
\]
The reason is that both the modes are equally distributed over two waveguides and the mode profiles inside 
the waveguides are considered to be identical due to small waveguide thickness. The equations (\ref{eqsX}) read
\begin{subequations}
\label{inteq}
\begin{eqnarray}
\frac{dP_S}{d\zeta}=\frac{2\kappa_0^{\prime\prime}P_S\Gamma_S}C\left(1+P_S\Gamma_S-P_A\Gamma_A\cos 2\phi\right),
\label{eq11}\\
\frac{dP_A}{d\zeta}=\frac{2\kappa_0^{\prime\prime}P_A\Gamma_A}C\left(1+P_A\Gamma_A-P_S\Gamma_S\cos 2\phi\right),
\label{eq12}\\
\frac{d\phi}{d\zeta}\approx L_R\delta\beta.
\label{eq13}\end{eqnarray}
\end{subequations}

\begin{figure*}
\includegraphics{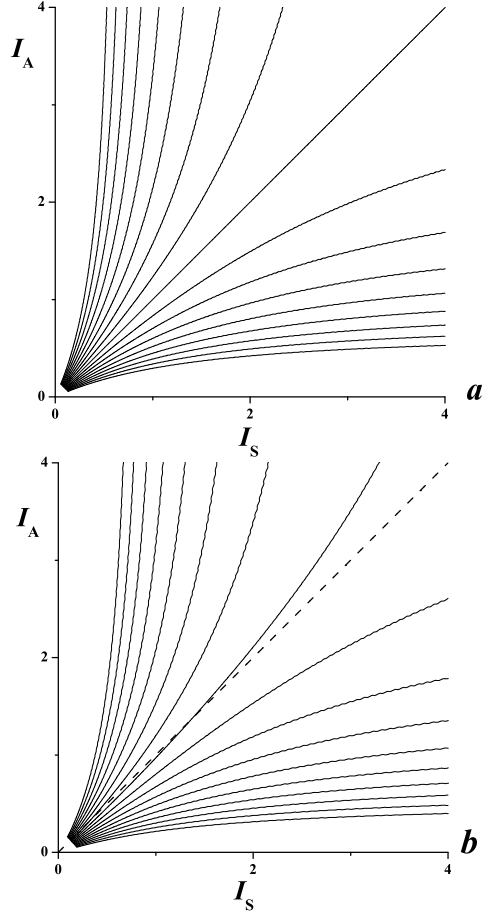}
\caption{\label{fig:chf}Mode intensities in the waveguide normalized to $I_{\text{Sat}}$  for different values of initial powers ratio: a) $\Gamma_S=\Gamma_A;$   b) $\Gamma_A=1.112\Gamma_S.$ Dashed line in fig. 4(b) is bisector 
$I_A=I_S.$}
\end{figure*}

Direct integration of (\ref{inteq}) was performed  by a Mathcad software solver based on $4^{\text{th}}$ order 
Runge-Kutta method. The calculations were made for the cases: 
a) $\kappa^{\prime}=0.900126$  $(\Gamma_A=\Gamma_S),$ and b) $\kappa^{\prime}=1.2$  $(\Gamma_A=1.112\Gamma_S).$ It was taken for calculations $\kappa_0^{\prime\prime}=kn_0g_0da/2=0.001.$ This value 
can be achieved at the small signal gain $g_0\approx 0.21~\text{cm}^{-1}$ for $a=5.2~\mu\text{m}$ and the 
other parameters of the construction as at the end of section \ref{sec4}, so that $\kappa^{\prime}=1.2$. The spectral separation of the modes in this case is $\delta\beta=0.277L_R^{-1}\approx 22~\text{cm}^{-1}.$   
 The results of calculations are presented in Fig. \ref{fig:chf}. 
For the equal confinement factors $\Gamma_A=\Gamma_S$ behavior of curves on diagram in fig. 4a showing $I_A$ as a
function of $I_S$ for different initial conditions is in contrast with their behavior in the case of incoherent 
modes (Fig. \ref{fig:inch}). In the last case, both modes increase power so as the proportion of powers is constant. In the 
former case, there is the line $I_S=I_A,$ which is unstable in Lyapunov's sense: an infinitesimal deviation from this trajectory results in further running away curves approaching asymptotically either vertical or horizontal line in dependence on direction of initial deviation. This behavior corresponds to dominance of one mode, power of which grows linearly with the length, while the power carried by the other mode is stabilized at a certain level. This behavior is similar to one observed in the Lamb's model \cite{Lamb} for two-mode laser in the case when the gain cross-saturation coefficient is greater than the self-saturation coefficient. 

If the confinement factors are different (see Fig. 4b), the diagram on the whole is nearly the same. 
The straight line $I_S=I_A$ in the diagram $I_A(I_S)$ in fig. 4a transforms to a curve (not shown in Fig. 4b),
the shape of which could be found numerically. This curve is a separatrix dividing $(I_S,I_A)$ plane into two 
parts. In the upper part all curves approach vertical asymptotes (asymmetric mode dominates), while in the 
lower part all curves approach horizontal asymptotes (symmetric mode dominates).

\begin{figure*}
\includegraphics{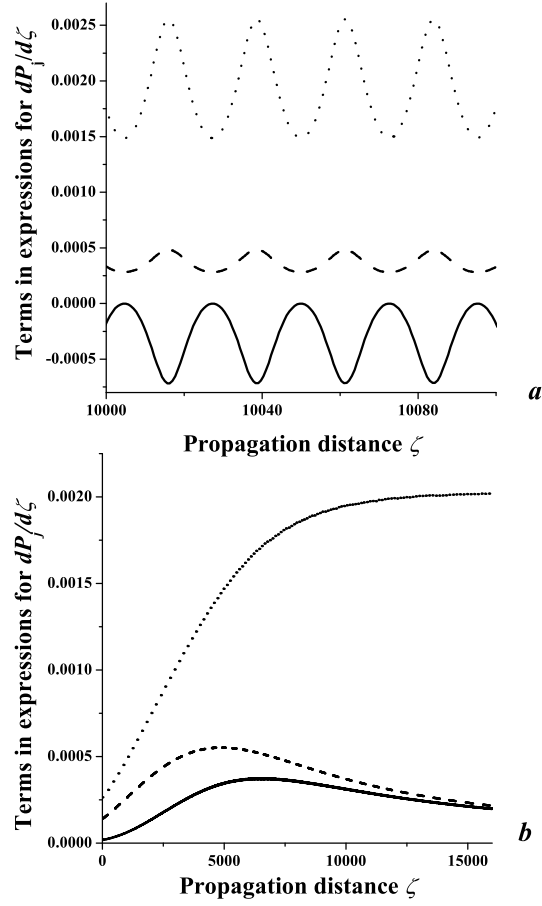}
\caption{\label{fig:det} Dimensionless terms $g_{AS}\cos\phi\sqrt{P_AP_S}L_R$(solid line),  $g_{AA}P_AL_R$ (dashed line), and $g_{SS}P_SL_R$ (dotted line) as functions of propagation distance normalized to $L_R$: (a) axial oscillations shown within a short propagation interval; (b) averaged over oscillations values,  $g_{AS}\cos\phi\sqrt{P_AP_S}L_R$(solid line) is taken with minus sign. $P_{S0}=0.27$ and $P_{A0}=0.13$; $\Gamma_A=1.112\Gamma_S.$ }
\end{figure*}

To get better insight into mechanism leading to depression one of the modes in two-mode amplification let us illustrate the general arguments adduced in section \ref{sec3} by results of analysis of the specific construction under consideration. 
Figure \ref{fig:det} illustrates behavior of the terms in the right hand side of eqs. (\ref{eqsX}) $g_{jj}P_j$ 
$\left(j=A,S\right)$ and $\sqrt{P_AP_S}g_{AS}\cos\phi$ with modal gain and cross gain coefficients defined by 
(\ref{coefcalc}). $P_{S0}=0.27$ and $P_{A0}=0.13.$ Small-scale oscillations of these terms associated 
with mode beating are shown in Fig. 5a. It is seen that antisymmetric mode power increase $g_{AA}P_A$ 
is lower than that of symmetric mode despite that its modal gain is higher. The cross gain component 
$\sqrt{P_AP_S}g_{AS}\cos\phi$ is non-positive and antiphase to the power increments associated with modal stimulated emission. The same terms averaged over oscillations are shown in Fig. 5b. For convenience of presentation the cross gain-term is multiplied by $(-1).$ It is clearly seen in Fig. 5b that decrement in emitted power caused by 
the cross-gain term tends to equilibrate the term  $g_{AA}P_A$ at a sufficiently long amplification length. 
In dimensional variables the Rayleigh length at parameters taken is $123~\mu\text{m}.$
This phenomenon leads to stabilization of the antisymmetric mode power on the level reached to that moment.

\begin{figure*}
\includegraphics{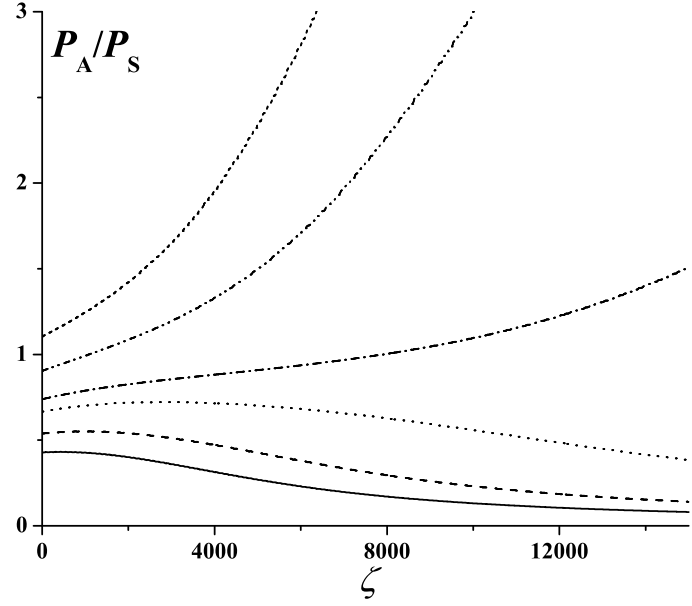}
\caption{\label{fig:prop}Proportion of the modal powers for varied inputs at constant total input power as a function of the amplifier length.  The quantities are dimensionless, $\Gamma_A=1.112\Gamma_S.$}
\end{figure*}

The trend to domination of one of modes in simultaneous amplification of two coherent modes can be illustrated 
by behavior of the mode power ratio $P_A/P_S.$ The proportion $P_A/P_S$ calculated for $\kappa^{\prime}=1.2$
is shown in Fig. \ref{fig:prop} as a function of propagation distance for various input proportions at the total input power kept constant $P_{A0}+P_{S0}=0.4$. is seen that increasing $P_{A0}/P_{S0}$ results in change of amplification 
regime from dominance of symmetric mode to asymmetric one. It is worth to note that the sign of derivative at
$\zeta=0$ of the proportion $P_A/P_S$  cannot serve as a criterion for the change of the amplification regime. 
As far as $P_{A0}/P_{S0}$ grows, the curve appears, for which growth at small distances changes to falling down 
at longer distances. Figure \ref{fig:prop} proves that there exists a critical value of  $P_{A0}/P_{S0},$ which separates regimes of dominance of symmetric or asymmetric mode. Actually, the critical value of $P_{A0}/P_{S0}$ is a function of the 
total power and depends parametrically on confinement factors values.

\begin{figure*}
\includegraphics{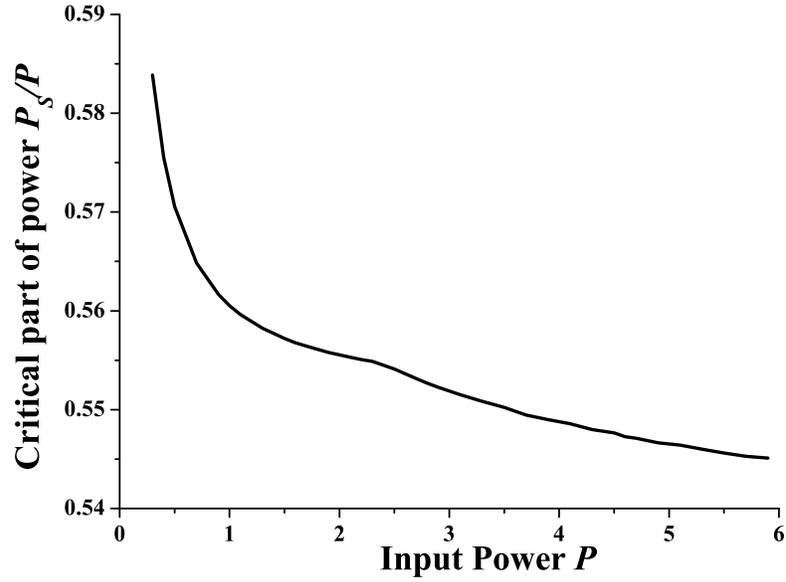}
\caption{\label{fig:frac} Critical fraction of the in-phase mode in the input total power $P=P_{A0}+P_{S0}$, 
above which this mode dominates.  The quantities are dimensionless, $\Gamma_A=1.112\Gamma_S.$}
\end{figure*}

A series of calculations allows us to find the critical fraction of the symmetric (in-phase) mode in the launched signal shown in Fig. \ref{fig:frac} as a function of the total power for confinement parameters values $(\Gamma_A=1.112\Gamma_S).$
At such values of $\Gamma_A$ and $\Gamma_S$ the small signal gain of the antisymmetric mode is greater than of the symmetric one. Despite this fact, the in-phase mode dominates at the output of an amplifier when its fraction in 
the total launched power is above curve shown in Fig. \ref{fig:frac}. The higher is the total launched power the less is 
the critical fraction of the in-phase mode. It follows from this figure that 40\% excess of the in-phase mode 
power in the input signal is sufficient to suppress the out-of-phase mode power for $P>0.3$
The analysis performed is strictly valid for the model system under consideration. However, the mechanism of weak mode suppression by a strong mode is of quite general nature. Thus, it is expected that this mechanism works in any fiber amplifiers provided the input signal has a sufficiently narrow spectral width in order that results found for monochromatic wave field were applicable for a real signal.

\section{\label{sec6}Conclusions}
Theory of monochromatic wave field amplification in a waveguide array is developed. An approach based on expansion 
of the wave field in terms of guided array modes leads to appearance of new terms in a system of ordinary evolution equation for the mode amplitudes. These terms have meaning of cross-modal gain and, as shown, completely change 
behavior of the amplified wave field. Analysis of two-mode amplification reveals new unusual features in 
characteristics of two-mode amplifier. Instead of unlimited growth of both modes for incoherent fields, 
the effect of weak-mode suppression by a strong one takes place. 
Detailed analysis is made for the amplifier composed of a pair of ultra-thin waveguides. The critical 
values of waveguides and input signal parameters are found, at which the in-phase mode dominate at 
the amplifier output. The conditions for asymptotically stable single-mode amplification are found. 
The model developed is applicable for studies on multimode competition in the multi-core fiber amplifier.

\begin{acknowledgments}
The work was partially supported by RFBR projects no. 07-02-01112-a, 07-02-12166-ofi
\end{acknowledgments}

\bibliography{FLA2}

\end{document}